\documentclass[10pt,sigconf,letterpaper]{acmart}

\usepackage{graphicx}
\usepackage{subcaption}
\usepackage{xcolor}
\usepackage{url}
\usepackage{hyperref}
\usepackage{cleveref}
\usepackage{xspace}
\usepackage{booktabs}
\usepackage[colorinlistoftodos]{todonotes} 

\AtBeginDocument{%
  \providecommand\BibTeX{{%
    \normalfont B\kern-0.5em{\scshape i\kern-0.25em b}\kern-0.8em\TeX}}}

\renewcommand\footnotetextcopyrightpermission[1]{}
\setcopyright{none}

\newcommand{\ie}{\emph{i.e.}\xspace}
\newcommand{\eg}{\emph{e.g.}\xspace}
\newcommand{\etal}{\emph{et al.}\xspace}

\begin{document}

\title[]{Internet Localization of Multi-Party Relay Users: \\ \huge Inherent Friction Between Internet Services and User Privacy}

\author{Sean Flynn}
\affiliation{%
  \institution{University of Hawai`i at M\=anoa}
  \city{}
  \country{}}
\email{}

\author{Francesco Bronzino}
\affiliation{%
  \institution{École Normale Supérieure de Lyon}
  \city{}
  \country{}
}
\author{Paul Schmitt}
\affiliation{%
  \institution{University of Hawai`i at M\=anoa}
  \city{}
  \country{}
}

\renewcommand{\shortauthors}{}

\sloppy

\begin{abstract}
  Internet privacy is increasingly important on the modern Internet. Users are
looking to control the trail of data that they leave behind on the systems that
they interact with. Multi-Party Relay (MPR) architectures lower the traditional
barriers to adoption of privacy enhancing technologies on the Internet. MPRs are
unique from legacy architectures in that they are able to offer privacy
guarantees without paying significant performance penalties. Apple's iCloud
Private Relay is a recently deployed MPR service, creating the potential for
widespread consumer adoption of the architecture. However, many current
Internet-scale systems are designed based on assumptions that may no longer hold
for users of privacy enhancing systems like Private Relay. There are inherent
tensions between systems that rely on data about users---estimated location of a
user based on their IP address, for example---and the trend towards a more
private Internet.

This work studies a core function that is widely used to control network and
application behavior, IP geolocation, in the context of iCloud Private Relay
usage. We study the location accuracy of popular IP geolocation services
compared against the published location dataset that Apple publicly releases to
explicitly aid in geolocating PR users. We characterize geolocation service
performance across a number of dimensions, including different countries, IP
version, infrastructure provider, and time. Our findings lead us to conclude
that existing approaches to IP geolocation (\eg, frequently updated databases)
perform inadequately for users of the MPR architecture. For example, we find
median location errors \textgreater1,000 miles in some countries for IPv4
addresses using IP2Location. Our findings lead us to conclude that new,
privacy-focused, techniques for inferring user location may be required as
privacy becomes a default user expectation on the Internet. 
\end{abstract}

\maketitle
\pagestyle{plain}

\section{Introduction}\label{sec:intro}
Nearly everything we do on the Internet leaves a trace, and in recent decades
the value of user data has proven to be highly-profitable and become a
fundamental business strategy of the Internet~\cite{schneier2015data,
zuboff2015big}. The only recourse users have for this situation is through
increased privacy, and it has long been understood that privacy is not solely a
benefit for individuals, but also aids society as a
whole~\cite{schmitt2022decoupling, gilliom2001overseers, mackinnon2012consent,
mcchesney2013digital, schneier2012liars, schneier2018}. It stands to reason,
then, that improving Internet privacy is a critical need, yet it is uniquely
difficult on the Internet because we must rely on others to carry and serve our
traffic.

The networking and security communities have long recognized the importance of
privacy, and have been designing solutions across the network stack for decades.
Beyond confidentiality (\ie, TLS encryption), which has thankfully been largely
solved, privacy research has focused on unlinkability between user identifiers
and their traffic. For example, even with encrypted payloads, the IP endpoints
of TCP sessions can be used to infer ``who is talking to whom.'' Chaum's
foundational mixnet work introduced the first architecture for anonymous
communication over the Internet~\cite{chaum1981untraceable}. Syverson \etal{}
furthered the mixnet concept and attempted to balance privacy and performance
with Onion Routing~\cite{syverson1997anonymous} and eventually
Tor~\cite{syverson2004tor}. While these systems have been widely available for
decades, their adoption has largely been limited to a small fraction of total
Internet users~\cite{2023_tor_metrics}. One potential reason for limited
adoption has been relatively poor and/or unpredictable
performance~\cite{dingledine2009performance, snader2008tune}.

Modern systems have sought to enhance user privacy without sacrificing
performance by leveraging the Multi-Party Relay
architecture~\cite{schmitt2022decoupling}. Apple's iCloud Private Relay (PR) is
a prime example of an MPR service. PR, currently in public beta~\cite{aboutpr},
was introduced by Apple in 2021 and the service is available to iCloud
subscribers. In PR, users pass through a first hop relay that is operated by
Apple and a second hop relay that is operated by one of Apple's infrastructure
partners (Akamai, Cloudflare, and Fastly) before egressing onto the open
Internet. Using the MPR architecture, Private Relay masks user IP addresses from
Internet hosts (as well as the second hop relay), while minimizing latency
penalties due to the addition of multiple middleboxes. In some ways, PR is
similar to a VPN but it offers improved privacy guarantees as VPN users must
trust the VPN provider, who is able to see both the user's identity and their
traffic. In contrast, the PR architecture ensures that neither Apple nor their
partners have access to both the user's identity and their traffic. The
integration of PR into Apple's platforms lowers the barriers to widespread
adoption of privacy-enhancing architectures, and has the opportunity to
influence future privacy-focused Internet systems.

The trend towards privacy brings inherent tensions with the practice of
leveraging information about the user to tailor network behavior---a common
example of this being IP geolocation. IP geolocation is a long-studied
networking problem as mapping IP addresses to location has many uses, including
routing client traffic to ``nearby'' CDN infrastructure, language localization,
and restricting content based on geography (\eg, live sports or other licensed
streaming content). Additionally, governments use geolocation data for emergency
planning, taxation, and regulation. IP geolocation is a fundamental component of
the modern Internet, yet it may be imperiled by the (justifiable) trend towards
enhanced user privacy. As users are decreasingly identifiable on the open
Internet, how will existing services perform? In this work, we study the
performance of popular IP geolocation services in the context of Apple's iCloud
Private Relay - a popular consumer MPR service. 

We conduct a measurement campaign to investigate the accuracy of popular IP
geolocation services with regard to Private Relay. PR is a unique service in
that it seeks to provide a specific dimension of privacy (\ie, masking user IP
addresses) while maintaining the expected user experience while using the
Internet. To accomplish this, Apple publishes fine-grained location information
for all of the IP prefixes that PR users are assigned to while using the system.
We compare geolocation service performance against the publicly available
location information for PR IP prefixes over the course of a 5 month period. We
analyze geolocation errors across a number of factors, including country,
infrastructure partner, IP geolocation service, IP version, and time. We find
that IP geolocation accuracy varies dramatically depending on these underlying
factors, and that geolocation errors for Private Relay users can be up to
thousands of miles.

Ultimately, our main goal in this work is not to solely focus on the accuracy
performance of IP geolocation services for iCloud Private Relay users. Instead,
we seek to to shed an initial light on inherent tensions between the status quo
of a core component (in this case localization) of the modern Internet and the
trend of increased user privacy. Certainly, localization of Internet users is
only one example of leveraging user metadata to personalize services, yet it
provides a tractable dimension for measurement and reasoning about system
performance as privacy is increased. Our findings lead us to conclude that
existing solutions are not currently viable for MPR users in cases where highly
accurate location information is required, and that fundamentally new designs
will be needed to provide equivalent localization functionality for users of
privacy-enhancing systems moving forward.

The rest of the paper is structured as follows: We first introduce PR and
discuss the context of our research in Section~\ref{sec:background}.
Section~\ref{sec:method} describes our proposed method. In
Section~\ref{sec:results} we present the results of our study. We discuss the
implication of our results in Section~\ref{sec:discussion}. Finally, we discuss
related work and conclude our study in Sections~\ref{sec:related}
and~\ref{sec:conclusion} respectively.

\section{Background}\label{sec:background}
In this section we provide a brief overview of the iCloud Private Relay
architecture. Further, we discuss the role of IP geolocation in the modern
Internet and the inherent challenges of providing geolocalization over privacy
focused architectures.

\begin{figure}[!t]
  \begin{minipage}{.85\columnwidth}
    \centering
    \includegraphics[width=1\columnwidth]{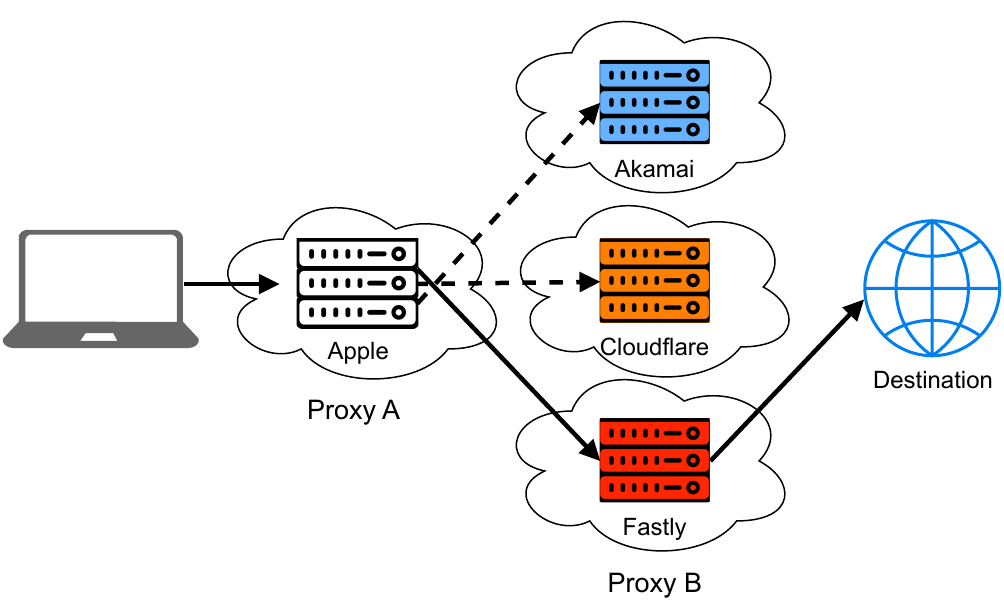}
    \caption{Private Relay Architecture}
    \Description{Private Relay Architecture}
    \label{fig:arch}
  \end{minipage}
\end{figure}

\subsection{Apple iCloud Private Relay}\label{sec:private_relay} Private Relay
(PR) aims to provide network-layer identifier (IP address) privacy of its users
by routing their Internet traffic through a Multi-Party Relay architecture.
Figure~\ref{fig:arch} shows a basic overview of the PR design. With PR, the
user's IP address is known to the first proxy (Proxy A in
Figure~\ref{fig:arch}), operated by Apple, but their traffic is not known as it
is encapsulated within an encrypted stream. The second proxy (Proxy B in
Figure~\ref{fig:arch}), operated by one of Apple's infrastructure partners
(Cloudflare, Fastly, and Akamai), is unable to ``see'' the user's IP address,
rather it only sees the IP address of the first proxy. However, the second proxy
may learn information about the user's request, such as the FQDN of the origin
server. Lastly, the origin server only learns the user's request and sees
traffic originating from the second proxy IP address. PR is available to all
Apple device users who have a subscription to iCloud+. Apple has incorporated PR
into its operating systems such as iOS, iPadOS, and
macOS~\cite{apple2021private}.

The PR architecture is similar in spirit to classic systems such as Chaum's
mixnets~\cite{chaum1981untraceable} or Tor~\cite{dingledine2004tor}, but differs
in two ways. First, it utilizes well-known rather than specialized protocols.
The default protocol for communication between the client and Proxy A is QUIC,
but if there are issues with QUIC setup, HTTP/2 and TLS is used as a fallback.
When communicating with Proxy B, HTTP/3 and Multiplexed Application Substrate
over QUIC Encryption (MASQUE)~\cite{schinazi2022http, schinazi2022rfc} are used.
If HTTP/3 is not available, the traditional HTTP CONNECT protocol over TLS is
used for communication with the second proxy~\cite{measuringpr}. Second, PR is
operated on commercial infrastructure rather than volunteer-run nodes and
includes functionality designed to maximize performance. For example, Cloudflare
utilizes Argo~\cite{lalkaka2017introducing}, a virtual network infrastructure
designed to optimize routing to minimize disruptions caused by
congestion~\cite{cloudflare2022icloud}. These strategies of the CDN egress
providers help to mitigate performance issues caused by introducing the
additional network hops necessary for an MPR
architecture~\cite{sattler2022towards}.

\subsection{IP Geolocation}\label{sec:ip_geolocation} IP geolocation involves
solving the problem of mapping the geographic location of a given IP address.
Many applications can benefit from using IP geolocation to determine the
geographic location of hosts on the Internet. For example, online advertisers
and search engines tailor their content based on the client's location. Content
providers can use IP geolocation to enforce digital rights management policies
and prevent users from accessing content that is not licensed for their
geographic location. In recent years, IP geolocation has become increasingly
relevant as Internet services aim to deploy low latency applications such as
cloud based online gaming and augmented reality.

Depending on the different use cases listed above, IP geolocation services can
be delivered with different degrees of granularity, varying from country level
localization down to city level. There exist two main approaches to geolocation
based on different techniques they employ: either relying on active network
measurements to determine the location of the host associated to the target IP
address or by querying an IP to geolocation mapping database. In this paper we
focus on the latter which is the common solution provided by commercial IP
geolocation services.

Typically, a geolocation database entry is composed of a pair of values,
corresponding to the integer representation of the minimum and maximum address
of an IP address block. Each block is then associated with several pieces of
information helpful for localization: country code, city, latitude and
longitude, and Zip code. There exist two classes of commercial databases: freely
accessible databases (\eg, MaxMind GeoLite2 City~\cite{mm}, IP2Location
LITE~\cite{ip2l}) and subscription based services (\eg, ipstack~\cite{ipstack}).
Free services are often advertised as slightly less accurate than their paid
counterparts.

\subsection{Challenges of IP Geolocation over Private Relay}\label{sec:challenges}

Any platform that aims to provide IP address privacy impacts the ability to map
a user's IP address to a geographic location. PR is unique in that it attempts
to find a balance between user privacy and user experience. Unlike VPNs, which,
in addition to masking user IP addresses are often used to provide a means to
access geographically filtered content\footnote{VPNs are also commonly used to
  circumvent Internet censorship. We could find no documentation that PR looks to
  provide censorship benefits~\cite{pr_censorship}.}~\cite{vpn_usage}, PR is
solely focused on masking users' IP addresses and it explicitly does not attempt
to provide these other functionalities. PR includes functionality in which Proxy
A establishes the client's location using traditional IP geolocation and then
provides a geohash to the client, which is subsequently offered to Proxy B in
order to map to a Proxy B IP address that is geographically near the client.

\begin{figure}[!t]
  \begin{minipage}{1\columnwidth}
    \centering
    \includegraphics[width=1\columnwidth,trim={60 100 60 40},clip]{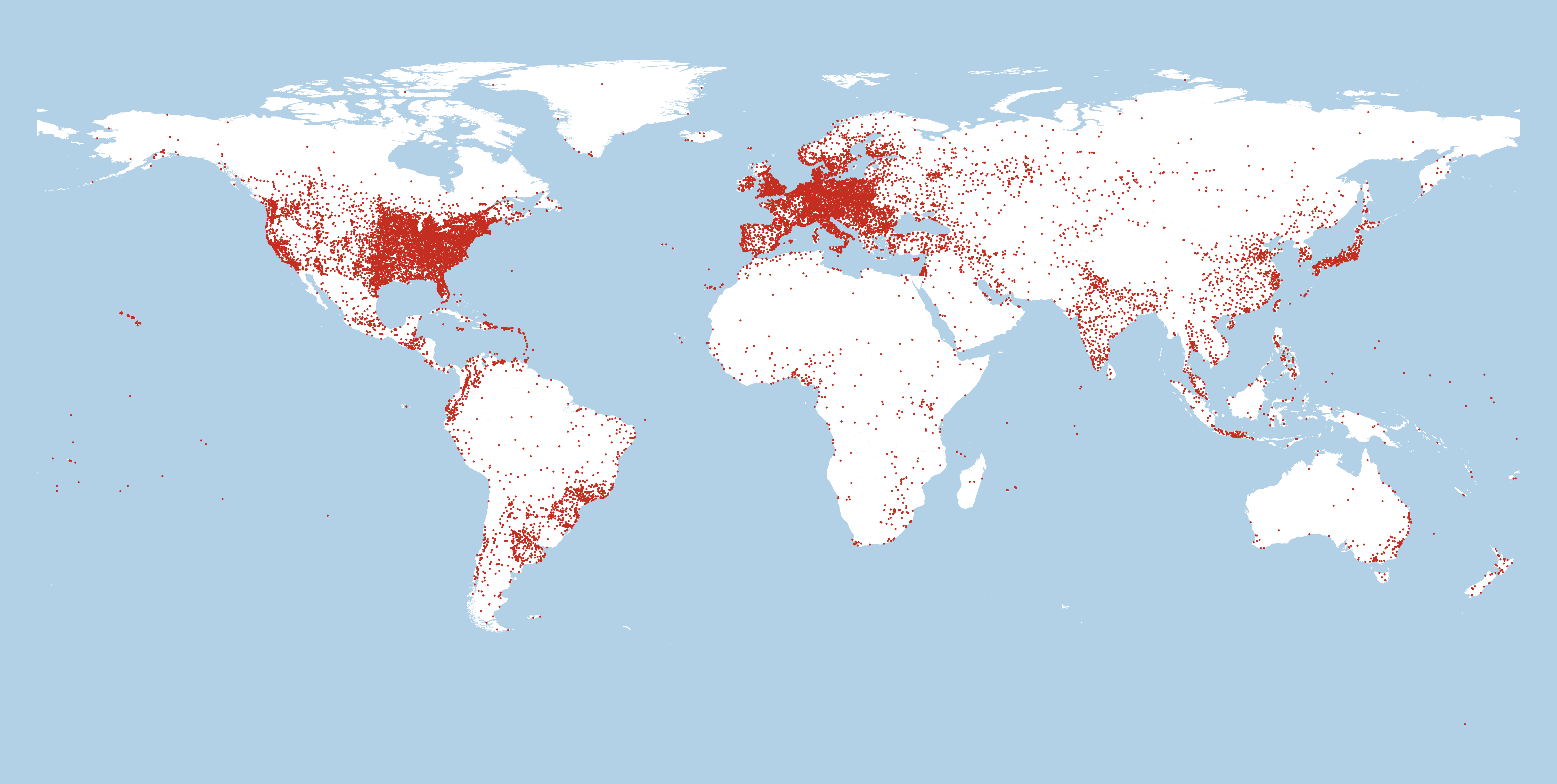}
    \caption{Advertised Apple iCloud Private Relay egress locations (red).}
    \Description{Advertised Apple iCloud Private Relay egress locations (red).}
    \label{fig:worldmap}
  \end{minipage}
  \hfill
\end{figure}

Apple and its infrastructure partners regularly publish Proxy B IP addresses and
their locations~\cite{egressips} in order to inform external services---\ie, IP
geolocation services---of ``accurate'' PR user locations while maintaining their
IP address privacy. Proxy B egress IP prefixes are advertised to be in locations
across a wide geographic footprint, as shown in Figure~\ref{fig:worldmap},
allowing for web services to localize PR users with relatively fine-grained
accuracy. Interestingly, the end result of the PR design is that advertised
egress IP address locations are, in many cases, obviously false. For example,
the egress IP dataset includes prefixes that are advertised to be hosted in
North Korea, even though we find no evidence of any of the three infrastructure
partners operating in North Korea. In a less extreme example, thousands of IP
prefixes are advertised in rural areas across the world.

\begin{table*}[h]

  \begin{tabular}{@{}l|rrrr|rrrr|r@{}}
                 & \multicolumn{4}{c|}{\textbf{IPv4}} & \multicolumn{4}{c|}{\textbf{IPv6}} &                                                                                                                                                                              \\ \cmidrule(lr){2-9}
    \textbf{CDN} & \multicolumn{1}{c}{US}             & \multicolumn{1}{c}{FR}             & \multicolumn{1}{c}{AU} & \multicolumn{1}{c|}{ZA} & \multicolumn{1}{c}{US} & \multicolumn{1}{c}{FR} & \multicolumn{1}{c}{AU} & \multicolumn{1}{c|}{ZA} & {\textbf{Total}}     \\ \midrule
    Akamai       & 28,816                             & 480                                & 594                    & 130                     & 267,846                & 6,720                  & 1,398                  & 812                     & 306,796              \\
    Cloudflare   & 6,030                              & 401                                & 290                    & 103                     & 46,684                 & 1,796                  & 1,240                  & 172                     & 56,716               \\
    Fastly       & 11,526                             & 110                                & 72                     & 8                       & 11,526                 & 110                    & 72                     & 8                       & 23,432               \\ \midrule
    Total        & 46,372                             & 991                                & 956                    & 241                     & 326,056                & 8,626                  & 2,710                  & 992                     & \multicolumn{1}{l}{}
  \end{tabular}
  \caption{Measured Private Relay Egress IPs by CDN and country on May 31, 2023.}
  \label{tab:egress}
\end{table*}

The focus of this work is to understand the interaction between these published
IPs and their locations versus the performance of popular IP geolocation
services to anticipate potential localization issues in reality.
\section{Methodology}\label{sec:method} In this section, we outline our data
collection methodology. We then describe the analysis we conduct and inherent
limitations.

\subsection{PR egress IPs and locations}
We establish ground truth using the egress (\ie, Proxy B in
Figure~\ref{fig:arch}) IP address information provided by Apple for the iCloud
Private Relay service~\cite{egressips}. The published file is a CSV that
includes fields for IP block, city, state (or region), and country. This dataset
is publicly released and periodically updated in order to allow for IP
geolocation services and or web services to associate Private Relay IP addresses
to the locations intended by Apple and their infrastructure partners. To assess
the evolution of PR egress IPs over time we download the file daily from January
1, 2023 through May 31, 2023.

\paragraph{Geographic ground truth.} As mentioned in
Section~\ref{sec:private_relay}, geographic ``ground truth'' is a somewhat
misleading term in Private Relay. In this work we assign ground truth to the
location that Apple and its infrastructure partners attest to, we do not attempt
to verify each location. We use Nominatim~\cite{nominatim}, an open source
geocoding tool based on OpenStreetMap, to establish baseline geographic
coordinates for each advertised egress location (\ie, geographic coordinates for
the center of a city listed in the published egress file). Note that while each
line in the file has fields for city, region, and country, not all records are
complete. Additionally, many locations in the file are listed under multiple
strings (\eg, Ciudad Juarez and Juarez, Mexico), resulting in an artificially
inflated number of unique locations. Overall, we observe a total of 18,110
unique location names in the dataset. Figure~\ref{fig:worldmap} shows a map with
the egress locations around the world. For brevity, we focus our analysis on
four countries: the United States, France, Australia, and South Africa. We
choose these countries as they represent a range of deployment and coverage
density, as shown in Table~\ref{tab:egress}.

\paragraph{IP address ground truth.}
We extract individual IP addresses out of their IP blocks differently depending
on the IP type. For IPv4 addresses, we include every individual address in our
measurements. Conversely, the IPv6 address prefixes in the dataset are typically
listed as /45 or /64 blocks, which include an infeasible number of individual
IPs to measure. Therefore, for each IPv6 prefix, we simply extract the first two
IP addresses for measurement purposes. We ran small experiments to test whether
random selection or larger numbers of IPv6 addresses would affect results, but
we observed no difference in performance. We also assign IPs to the respective
infrastructure partner by performing autonomous system \texttt{whois} lookups
for each prefix.

Table~\ref{tab:egress} shows PR egress IP addresses categorized by autonomous
system and IP version for each of Apple's three infrastructure partners on a single
day. As shown, Akamai offers the largest number of both IPv4 and IPv6 address
blocks, followed by Cloudflare and Fastly, respectively.

\begin{figure*}[t]
    \begin{minipage}{.245\textwidth}
        \centering
        \includegraphics[width=\textwidth]{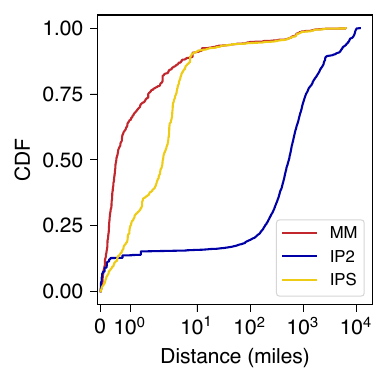}
        \caption{US IPv4}
        \Description{United States IPv4}
        \label{fig:us_v4}
    \end{minipage}
    \hfill
    \begin{minipage}{.245\textwidth}
        \centering
        \includegraphics[width=\textwidth]{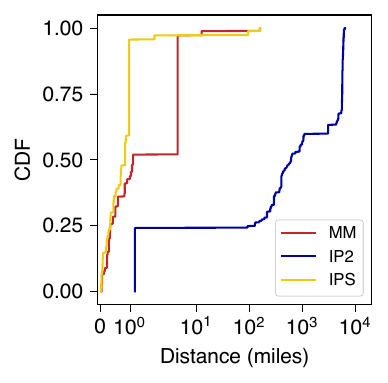}
        \caption{FR IPv4}
        \Description{France IPv4}
        \label{fig:fr_v4}
    \end{minipage}
    \hfill
    \begin{minipage}{.245\textwidth}
        \centering
        \includegraphics[width=\textwidth]{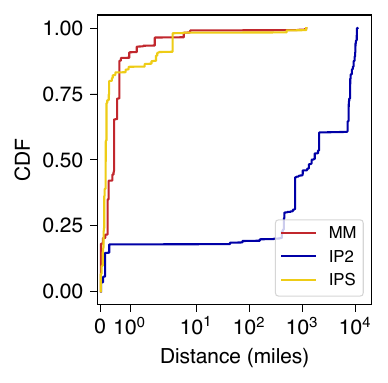}
        \caption{AU IPv4}
        \Description{Australia IPv4}
        \label{fig:au_v4}
    \end{minipage}
    \hfill
    \begin{minipage}{.245\textwidth}
        \centering
        \includegraphics[width=\textwidth]{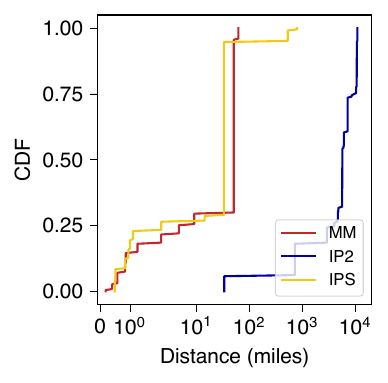}
        \caption{ZA IPv4}
        \Description{South Africa IPv4}
        \label{fig:za_v4}
    \end{minipage}

    \begin{minipage}{.245\textwidth}
        \centering
        \includegraphics[width=\textwidth]{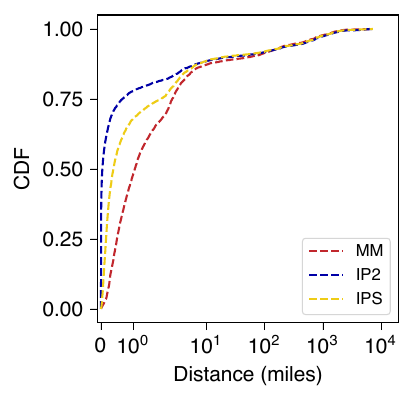}
        \caption{US IPv6}
        \Description{United States IPv6}
        \label{fig:us_v6}
    \end{minipage}
    \hfill
    \begin{minipage}{.245\textwidth}
        \centering
        \includegraphics[width=\textwidth]{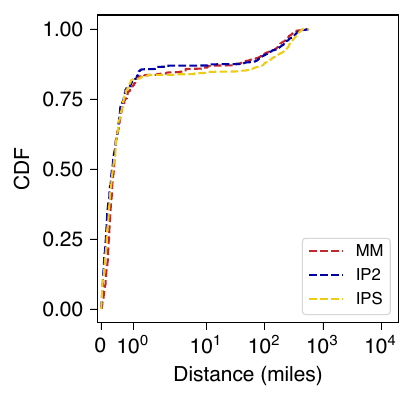}
        \caption{FR IPv6}
        \Description{France IPv4}
        \label{fig:fr_v6}
    \end{minipage}
    \hfill
    \begin{minipage}{.245\textwidth}
        \centering
        \includegraphics[width=\textwidth]{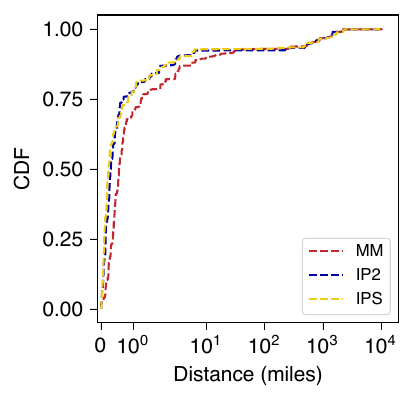}
        \caption{AU IPv6}
        \Description{Australia IPv4}
        \label{fig:au_v6}
    \end{minipage}
    \hfill
    \begin{minipage}{.245\textwidth}
        \centering
        \includegraphics[width=\textwidth]{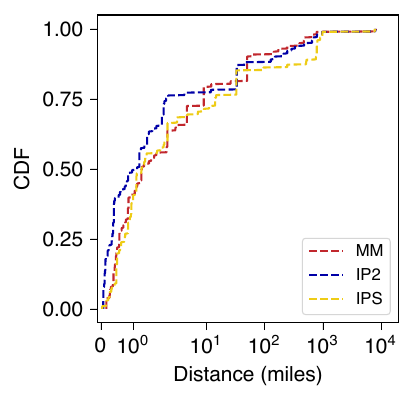}
        \caption{ZA IPv6}
        \Description{South Africa IPv6}
        \label{fig:za_v6}
    \end{minipage}
\end{figure*}

\subsection{Analyses}

\paragraph{IP Geolocation services.}
We use three popular IP geolocation services to compare computed locations based
on IP address versus the published ground truth dataset. The services we use are
MaxMind GeoLite2 City~\cite{mm}, IP2Location LITE~\cite{ip2l}, and
ipstack~\cite{ipstack}. The MaxMind and IP2Location services are commonly used
free services, while ipstack is a commercial API. Each of the services outputs
the geographic coordinates associated with an IP address at the city level. Web
services commonly use geolocation services to tailor (or often restrict) content
based on a user's location. We choose the city-level services as many streaming
services (\eg, live sports) require localizing users at a level finer than
country-level.

To understand temporal performance of the service, we explore MaxMind's
performance over time (Section~\ref{sec:mm_over_time}) by downloading every
release of the MaxMind database during the measurement campaign. MaxMind
publicly releases the database twice per week, and we download versions from
December 30, 2022 through May 30, 2023.

\paragraph{Geographic error.}
For each IP address, we calculate the distance between the ground truth
coordinates and the geolocated coordinates from each service, allowing us to
reason about how web services would localize users connecting from PR egress IP
addresses, and whether those locations match locations from which Apple intends
them to appear.

\paragraph{Limitations}
Our study is necessarily limited to the IP geolocation services that we have
selected. Certainly, other geolocation services may perform better or worse
compared with those studied. As this study is meant to shine a light on
potential web localization issues that could arise from privacy-preserving
system architectures, we leave in-depth, service-by-service exploration for
future work. We also do not study the relationship between the actual location
of PR users and the egress locations they are assigned to as this is proprietary
information. For simplicity, we assume the general use case of PR is that users
are assigned egress IPs and locations that are geographically closest to their
actual locations.
\section{Results}\label{sec:results}
In this section we analyze IP geolocation services with respect to PR across a
number of dimensions, including country, IP version, IP geolocation service,
PR infrastructure partner, and time.

\subsection{Country-wide IP geolocation service performance}\label{sec:per_country}

We begin by studying IP geolocation service accuracy for each country in our
dataset. For this experiment, we choose a single day (March 15, 2023) to
calculate geographic distance between the advertised PR egress IP locations and
the geolocated
coordinates.~\Cref{fig:us_v4,fig:us_v6,fig:fr_v4,fig:fr_v6,fig:au_v4,fig:au_v6,fig:za_v4,fig:za_v6}
display the results for IPv4 and IPv6 for each country and for each service
(MaxMind is labeled `MM', IP2Location is labeled `IP2', and ipstack is labeled
`IPS').

For IPv4 addresses, we observe that IP2Location (IP2) tends to perform worst
among the geolocation services across all of the countries studied, oftentimes
showing distance errors 5,000 - 10,000 miles larger that the other two services.
For every country, IP2 performs relatively well for a small percentage of the
IPv4 addresses; for example, roughly 20\% of Australian IPv4 errors are less
than 1 mile. However, the median distance error for IP2 in Australia is 1,479
miles. The other two services (MM and IPS) result in significantly lower
distance errors on the whole, with neither service performing particularly
better than the other. When looking from a country-wide perspective, we observe
the highest accuracy in Australia for MM and IPS, with both services showing
\textgreater90\% of distance errors less than 10 miles. We posit that Australia
may benefit from the nature of its population concentration in relatively few
areas along the coasts, whereas the populations of the other countries studied
are more evenly spread geographically. Australia's dense urban areas offer more
obvious locations for CDNs to install infrastructure, in comparison. The United
States and France both also show results that would appear acceptable for MM and
IPS, with median errors of less than 2 miles for the services in both countries.
As one might anticipate, South Africa, the country with the least dense PR
deployment (Table~\ref{tab:egress}), has the highest median errors for all three
services (50, 5,592, and 33 miles for MM, IP2, and IPS, respectively).

Interestingly, IPv6 results (\Cref{fig:us_v6,fig:fr_v6,fig:au_v6,fig:za_v6})
differ significantly compared to IPv4 addresses. We see that all geolocation
services tend to perform similarly for each country. IP2Location, the worst
performing service for IPv4 addresses, shows the highest accuracy for the United
States and South Africa, and comparative accuracy for France and Australia.
Additionally, IPv6 distance error distributions tend to be dominated by highly
accurate measurements (\ie, less than 10 miles) with a long tail reaching
thousands of miles. Overall, our results indicate that PR users that are given
IPv6 addresses are more likely to be accurately geolocated compared to users
assigned to IPv4 PR addresses.

\subsubsection{IPv6 distribution similarity}\label{sec:v6similarity} As shown
in~\Cref{fig:us_v4,fig:fr_v4,fig:au_v4,fig:za_v4}, the error distributions for
the IP geolocation services can differ widely for IPv4 egress addresses.
However, we notice that the error distributions for IPv6 addresses qualitatively
appear similar for many of the countries
(\Cref{fig:us_v6,fig:fr_v6,fig:au_v6,fig:za_v6}). To explore this
quantitatively, we perform one-way ANOVA analysis to discover whether the error
distributions are statistically similar. The output of one-way ANOVA is a
p-value, which can be understood to indicate whether the differences in the
means of multiple distributions are statistically significant. If the p-value is
less than or equal to $0.05$ we reject the null hypothesis that states that the
means are equal. Higher p-values suggest that we do not have sufficient evidence
to reject the null hypothesis, and that the distribution means could be equal.
Table~\ref{tab:anova} shows the results. We see that the p-value for all
countries is far below $0.05$ for IPv4, meaning we reject the null hypothesis.
Conversely, we observe significantly higher p-values for IPv6 and both Australia
and South Africa, meaning the three error distributions could be considered
similar.

\begin{table}[ht]
    \centering
    \begin{tabular}{@{}crr@{}}      & \multicolumn{2}{c}{One-way ANOVA p-value}                     \\
               \toprule Country & \multicolumn{1}{r}{IPv4}                  &
               \multicolumn{1}{r}{IPv6}                                                         \\ \midrule
               US               & $0.0$                                     & $1.28\text{e-}58$ \\
               FR               & $3.06\text{e-}251$                        & $6.41\text{e-}27$ \\
               AU               & $4.47\text{e-}290$                        & $0.967$           \\
               ZA               & $2.41\text{e-}159$                        & $0.155$           \\ \bottomrule
    \end{tabular}
    \caption{One-way ANOVA results. IPv6 error distributions in some countries could be argued to be similar (p-value \textgreater 0.05) across geolocation services, while IPv4 error distributions are not similar (p-value \textless 0.05).}
    \label{tab:anova}
\end{table}

The precise reason for these results is unclear. We posit that the IP
geolocation services may have added new IPv6 entries as a direct result of Apple
publicly releasing the ground truth egress location dataset - meaning all of the
services would be working from the same, relatively recent starting point.
Conversely, all or nearly all IPv4 addresses have been present in IP geolocation
databases for several years and could have changed ownership or
location---particularly IP addresses associated with public cloud infrastructure
can be ``moved'' readily. Unfortunately, the IP geolocation services that we
leverage in this work do not publish historical information about IPs in their
databases. Further investigation into the discrepancy in performance between
IPv4 and IPv6 addresses is left to future work.

\subsubsection{Characterizing large errors}\label{sec:huge_errors}

We look to further understand extremely large distance errors that we encounter
when processing the dataset. For each record in the countries we study with a
distance error \textgreater1,000 miles, we compare the ground truth country
information versus the geolocated country along with the infrastructure partner
that owns the IP prefix.

We find that behavior varies between different combinations of geolocation
service, infrastructure, and country. MaxMind returns the correct country for
all large errors in both Australia and South Africa, but returns a number of
different (incorrect) countries for IP prefixes in France and the United States.
For French IP prefixes that are operated by Fastly, MaxMind geolocates to
Belgium, and, more interestingly, some in Uruguay. Conversely, French prefixes
operated by Akamai are geolocated to France. MaxMind is most interesting in the
United States--while the vast majority of large geolocation errors are
geolocated within the United States, a small fraction return locations in 32
different countries, including India and Guatemala for Fastly-owned prefixes,
and Afghanistan for Akamai-owned prefixes. Notably, we find that
\textgreater1,000 mile errors for MaxMind do not occur for Cloudflare prefixes
in any country outside of Australia.

IP2Location, generally the least accurate geolocation service, expectedly
results in quite different behavior for large errors. In Australia, the majority
(89.9\%) of \textgreater1,000 mile errors are associated with Cloudflare IP
prefixes. Akamai prefixes never result in large errors, while Fastly accounts
for \textasciitilde10\%. Australian prefixes are most commonly geolocated to the
United States for Cloudflare, and to Sweden for Fastly IPs when using
IP2Location. South Africa and France demonstrate roughly similar behavior, with
erroneous Cloudflare prefixes most commonly associated to the United States and
Fastly prefixes to Sweden. This behavior appears be explained by \texttt{whois}
information associated with the prefixes owned by the infrastructure partners. A
large number of prefixes owned by Fastly are registered as being located in
Sweden, while Cloudflare's prefixes are associated with their headquarters in
the United States. We believe that IP2Location may fall back to basic
\texttt{whois} information for an IP address if the database does not include
more specific location information for the associated IP prefix.

Large errors with IP2Location in the United States provide perhaps the most
interesting results. For example, a large number of errors for prefixes owned by
Akamai are geolocated to Australia. Fastly prefixes are again most often
erroneously mapped to Sweden, but they are additionally geolocated to Cyprus and
Macau. Cloudflare prefixes are geolocated to Japan and Great Britain. Overall,
we observe no discernable pattern when investigating large geolocation errors
for IP prefixes that are meant to be located in the United States. We believe
additional study into the nature of geolocation errors for core Internet
infrastructure IP addresses (\ie, rather than residential or edge addresses) is
warranted and we leave this for future work.

\subsubsection{False ground truth locations}\label{sec:north_korea}

As mentioned in Section~\ref{sec:private_relay}, the ground truth dataset
advertises many locations where there is almost certainly no partner cloud
infrastructure deployed. One example of this is North Korea. The May 31, 2023
egress dataset advertises 39 IP prefixes located in North Korea: 24 IPv4
prefixes and 15 IPv6 prefixes. The prefixes are hosted by Akamai and Cloudflare,
and we find no evidence of Akamai or Cloudflare deployments in North Korea.

We attempt to geolocate North Korean IP addresses in each of the prefixes using
all three services. Table~\ref{tab:kp} shows the country codes returned by each
of the services for IPv4 addresses. As shown, both MaxMind and ipstack, the
more accurate services in our prior analysis, report North Korea for all of the
IP addresses. IP2Location, on the other hand, never reports North Korea as the
country and instead often reports Japan as the location. We attempt to verify
the actual location of the IP addresses using looking glass servers located
around the world. We find evidence that many of the IP addresses are in fact
hosted on infrastructure in Japan (\ie, ping RTTs of \textasciitilde1-2ms from
looking glass servers located in Japan).

\begin{table}[h]
    \begin{tabular}{@{}l|rrrr|@{}}
                    & \multicolumn{4}{|c|}{Country}                \\ \midrule
                    & KP                            & JP & TW & US \\
        MaxMind     & 24                            & 0  & 0  & 0  \\
        IP2Location & 0                             & 18 & 1  & 5  \\
        IP Stack    & 24                            & 0  & 0  & 0  \\ \bottomrule
    \end{tabular}
    \caption{IPv4 country locations for IP prefixes advertised as within North Korea (KP).}
    \label{tab:kp}
\end{table}

Conversely, all three of the geolocation services report North Korea as the
country for every IPv6 address in the dataset. This result furthers the notion
that IP version has a dramatic impact on geolocation accuracy when using PR. As
seen in~\Cref{fig:us_v6,fig:fr_v6,fig:au_v6,fig:za_v6}, IPv6 errors tend to be
similar across all geolocation services.

\begin{figure*}[t]
    \begin{minipage}{.245\textwidth}
        \centering
        \includegraphics[width=\textwidth]{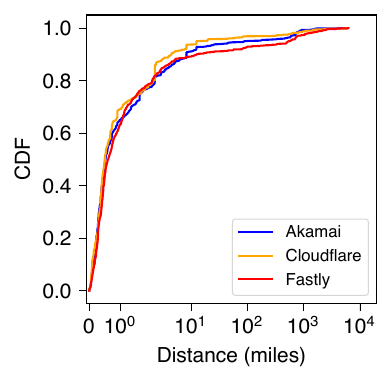}
        \caption{MM US}
        \Description{MaxMind United States}
        \label{fig:us_mm_cdn}
    \end{minipage}
    \hfill
    \begin{minipage}{.245\textwidth}
        \centering
        \includegraphics[width=\textwidth]{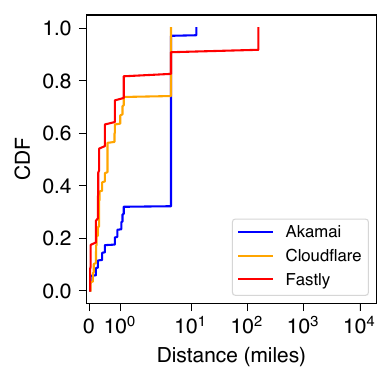}
        \caption{MM FR}
        \Description{MaxMind France}
        \label{fig:fr_mm_cdn}
    \end{minipage}
    \hfill
    \begin{minipage}{.245\textwidth}
        \centering
        \includegraphics[width=\textwidth]{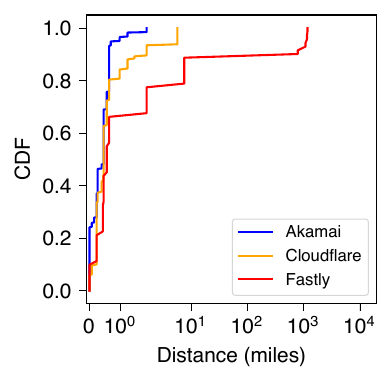}
        \caption{MM AU}
        \Description{MaxMind Australia}
        \label{fig:au_mm_cdn}
    \end{minipage}
    \hfill
    \begin{minipage}{.245\textwidth}
        \centering
        \includegraphics[width=\textwidth]{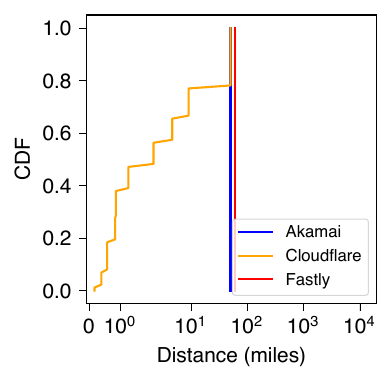}
        \caption{MM ZA}
        \Description{MaxMind South Africa}
        \label{fig:za_mm_cdn}
    \end{minipage}

    \begin{minipage}{.245\textwidth}
        \centering
        \includegraphics[width=\textwidth]{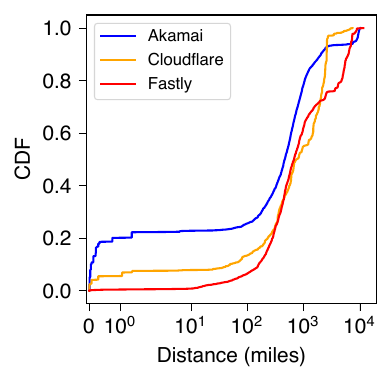}
        \caption{IP2 US}
        \Description{IP2Location United States}
        \label{fig:us_ip2_cdn}
    \end{minipage}
    \hfill
    \begin{minipage}{.245\textwidth}
        \centering
        \includegraphics[width=\textwidth]{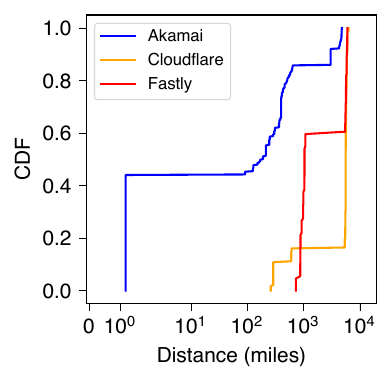}
        \caption{IP2 FR}
        \Description{IP2Location France}
        \label{fig:fr_ip2_cdn}
    \end{minipage}
    \hfill
    \begin{minipage}{.245\textwidth}
        \centering
        \includegraphics[width=\textwidth]{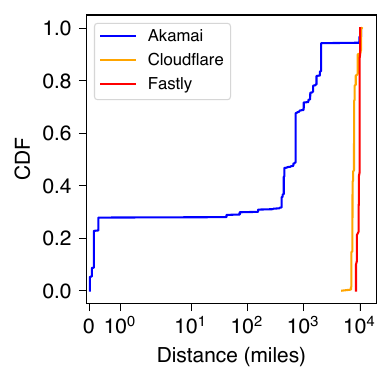}
        \caption{IP2 AU}
        \Description{IP2Location Australia}
        \label{fig:au_ip2_cdn}
    \end{minipage}
    \hfill
    \begin{minipage}{.245\textwidth}
        \centering
        \includegraphics[width=\textwidth]{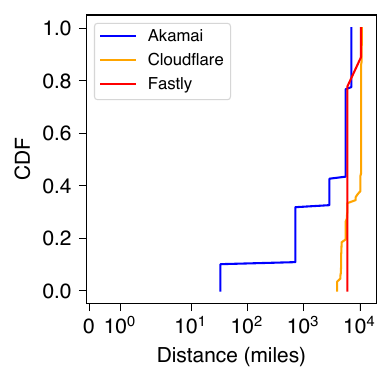}
        \caption{IP2 ZA}
        \Description{IP2Location South Africa}
        \label{fig:za_ip2_cdn}
    \end{minipage}
\end{figure*}

These results beg the question: what is geolocation ``accuracy'' for a system
that purposefully falsifies location information of its users? One could argue
that closely matching Apple's advertised location is the ideal outcome for web
services (\eg, the behavior of MaxMind and ipstack, in this case). Of course,
this assumes that PR users are mapped to the geographically closest egress IP
addresses when using the system. Alternatively, one could argue that IP2Location
is the more accurate service for these IP addresses, as it is more accurately
characterizing the locations in which user traffic actually ingresses and
egresses from the Internet. While this discussion is beyond the scope of a
measurement campaign, we briefly discuss the overall goals of IP geolocation in
Section~\ref{sec:discussion}.

\subsection{Infrastructure partner IP geolocation service performance}\label{sec:per_cdn}

Next, we investigate IPv4 geolocation service performance with respect to the
different infrastructure partners that operate PR egress infrastructure (Akamai,
Cloudflare, and Fastly). For brevity, and given that we found MaxMind and
ipstack to perform similarly in Section~\ref{sec:per_country}, we only focus on
results using MaxMind (MM) and IP2Location (IP2). We again calculate the
geographic error distance between the Apple published egress ground truth
locations and the geolocated coordinates, only we now categorize based on the
partner AS that owns the IPs in
question.~\Cref{fig:us_mm_cdn,fig:fr_mm_cdn,fig:au_mm_cdn,fig:za_mm_cdn} show
the results for MaxMind
and~\Cref{fig:us_ip2_cdn,fig:fr_ip2_cdn,fig:au_ip2_cdn,fig:za_ip2_cdn} show the
results for IP2Location.

We find that it appears as though different combinations of infrastructure
partner and IP geolocation service matters greatly in terms of accuracy. For
instance, Akamai IP addresses tend to be geolocated with the highest accuracy
when using IP2, but the same cannot be said for MM, with Akamai IP addresses
often resulting in the worst or nearly the worst accuracy. Curiously, both
Akamai and Fastly South African IP addresses result in consistent values when
using MaxMind. We investigate and find that they each repeatedly geolocate to
static locations and their corresponding ground truth locations are similar. For
instance, MaxMind ground truth IPv4 addresses are located in Pretoria, while the
geolocated coordinates are south of Johannesburg, 50.2 miles away.

We are unsure as to why there are significant localization performance gaps
depending on the infrastructure partner and IP geolocation service, given that
the ground truth location information is kept up to date and published by Apple
in a single file for all partners combined. We posit that the differences we
witness could be historical artifacts based on IP re-use practices of the
partners themselves or on the ground truth IP location ingestion practices of
the geolocation services (\eg, how each service manages partially complete
records in the ground truth dataset).

This analysis shows us that the combination of PR infrastructure partner and the
IP geolocation service used to locate a given user can have a significant impact
on the ability to accurately localize a user. Unfortunately, the method used to
map users to egress IPs (and thus to infrastructure partners) is not public
information. Likewise, the vast majority of services on the Internet that use IP
geolocation service do not advertise which service(s) that they subscribe to.
Ultimately, this means that users have little predictive control over
localization of themselves when using PR to connect to Internet services. This
result, along with the scale of the error for some of the services (\eg, medians
of 1,479 miles and 5,592 miles for IP2 in Australia and South Africa,
respectively), lead us to conclude that geolocalization outcomes for users of
systems like Private Relay are non-deterministic as of today.

\subsection{Temporal evolution of Private Relay egress IPs and geolocation services}\label{sec:temporal}

Next, we investigate the published egress IP and location dataset to understand
it's evolution over time. Our intuition is that a high churn rate of IP prefixes
or locations could negatively impact the accuracy of IP geolocation services.
For brevity, we limit geolocation accuracy results to MaxMind, the
best-performing geolocation service provider, for the remainder of this paper.
We then study the accuracy performance of MaxMind over time as the database is
updated to reflect changes in the PR egress dataset.

\subsubsection{Stability of egress IP addresses over time}
We first analyze the published egress dataset to highlight changes that occur
over the course of the observation period. As Private Relay is a relatively new
service, it stands to reason that the egress dataset is evolving over time.
Further, we anticipate that major changes to the dataset could negatively effect
the accuracy of IP geolocation services when locating PR egress IP addresses, as
lag time in database updates would manifest in location errors.

\begin{figure}[!t]
    \begin{minipage}{.48\textwidth}
        \includegraphics[width=\columnwidth]{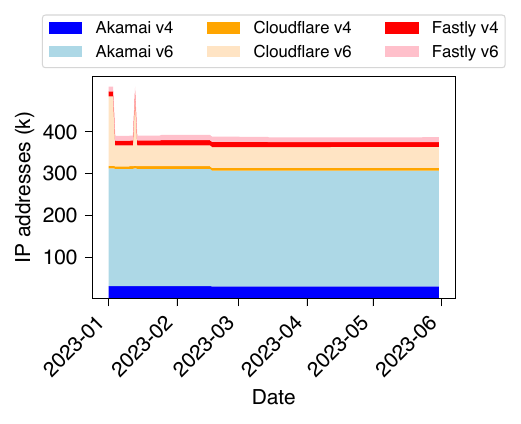}
        \caption{Measurable egress IP addresses advertised over time.}
        \Description{Measured egress IP addresses advertised over time.}
        \label{fig:ips_over_time}
    \end{minipage}%
\end{figure}

\paragraph{IP address space}
Figure~\ref{fig:ips_over_time} shows the measured IP addresses in the four
countries we study during the observation period categorized by infrastructure
partner and IP version. We see that in the beginning of the measurement
campaign, a large number of IPv6 addresses owned by Cloudflare we advertised and
subsequently removed before another short-lived increase roughly one week later.
The other infrastructure partners tend to advertise a more stable number of IP
addresses throughout the study. Akamai offers significantly more addresses than
Cloudflare or Fastly throughout. Overall,the total count of IP addresses does
not drastically change over the course of the observation period. It would
logically follow that short-lived IP addresses such as those offered by
Cloudflare in the beginning of the study could potentially experience poor
geolocation accuracy due to lag between the IP addresses being newly advertised
as part of the PR egress service and geolocation services updating their
databases. We explore this further in Section~\ref{sec:mm_over_time}.

\paragraph{Added, split, and merged IP prefixes}
While the total count of measurable IP addresses remains relatively stable in
the observed countries, further exploration reveals that the egress IP prefixes
evolve more subtly. We find that a significant number of IP prefixes are added
during the measurement campaign. Over the duration of our observation, we find
a total of 10,131 new IP prefixes were added representing 19,663 measurable IP
addresses - roughly equivalent to 5\% of the total IP address space. This result
is unintuitive given the stability shown in Figure~\ref{fig:ips_over_time}, but
can be explained by observing aggregation and disaggregation of IP prefixes into
different sized CIDR blocks. We observe many instances of IP prefixes being
split (\ie, disaggregated into smaller CIDR blocks) and merged (\ie, aggregated
into larger CIDR blocks) over time.

\begin{figure}[!t]
    \begin{minipage}{.45\textwidth}
        \centering
        \includegraphics[width=\columnwidth]{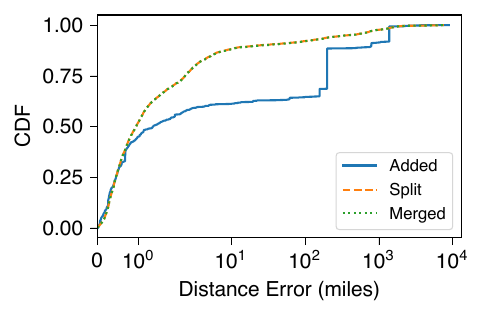}
        \caption{Added IP prefixes}
        \Description{Added IP prefixes}
        \label{fig:addsplitmerge}
    \end{minipage}
\end{figure}

Over the duration of our observation, 132,377 prefixes were split, and 23,713
prefixes were merged. We measure the accuracy performance of MaxMind for added, merged, and split IP prefix and plot the results in~\Cref{fig:addsplitmerge}. As
shown, newly added IP addresses tend to result in lower geolocation accuracy
compared with addresses belonging to split or merged IP prefixes. We also see
that split and merged IP prefixes result in nearly identical error
distributions. We find that split and merged prefixes were all owned by
Cloudflare, and we believe this accuracy performance could be attributable to
split or merged prefixes not changing location at the time of a change,
resulting in no change in geolocation accuracy.

\subsubsection{Stability of egress geographic locations over time}
In addition to newly added, split, and merged IP prefixes in the dataset, we
also observe a small number of prefixes whose location is changed in the ground
truth egress dataset. In this analysis we include all moved IP prefixes, not
limited to the four countries presented previously. We include worldwide moves
to illustrate the large changes that are present in the dataset.

\begin{figure}[t]
    \centering
    \begin{minipage}{1\columnwidth}
        \includegraphics[width=1\columnwidth,trim={40 200 40 100},clip]{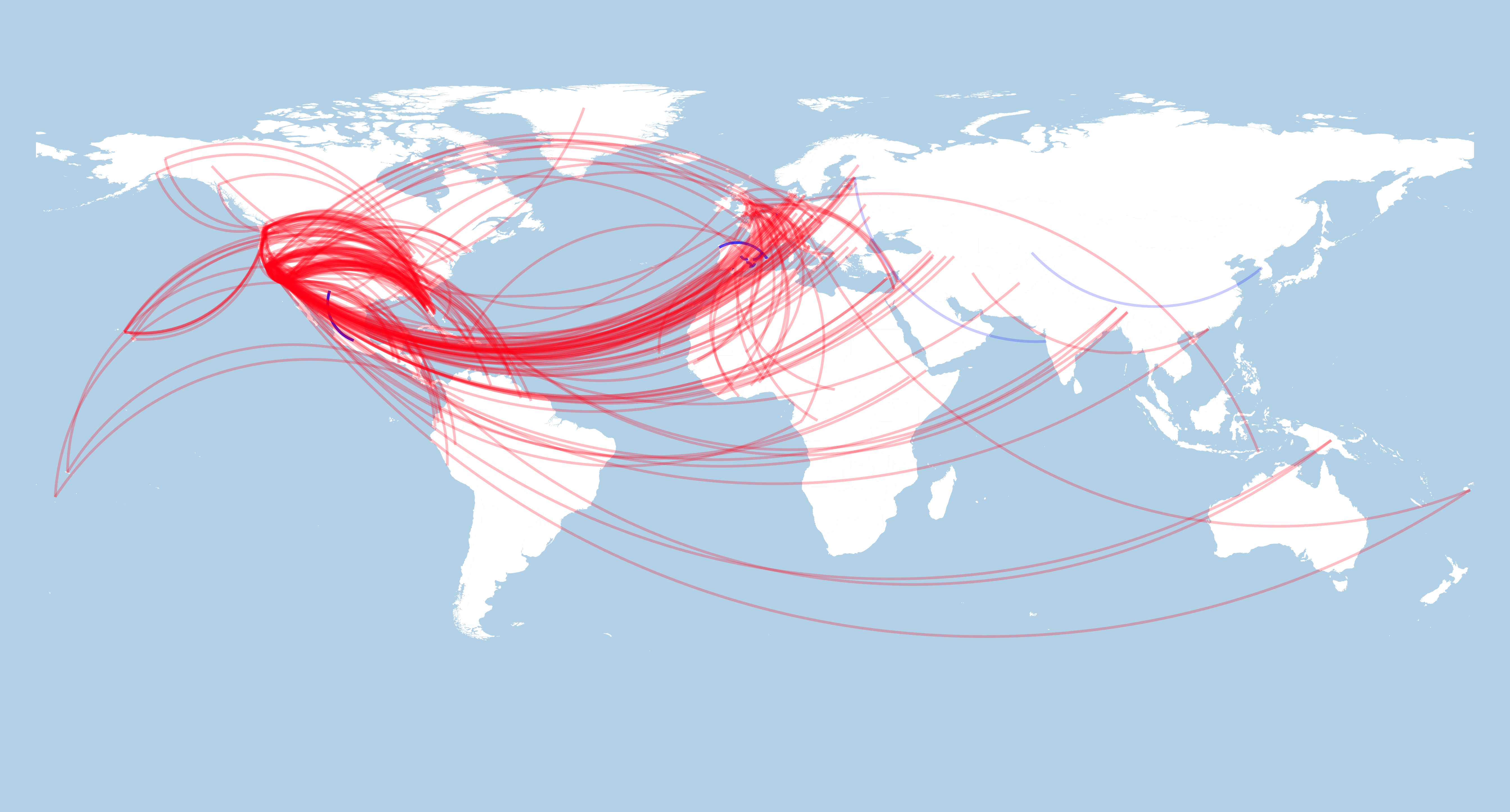}
        \caption{IP prefixes that moved during the observation period. Arcs indicate direction of the move (clockwise). Color indicates infrastructure partner: Akamai in blue, Fastly in red, Cloudflare did not have any moved IP prefixes.}
        \Description{IP prefixes that moved during the observation period. Arcs indicate direction of the move (clockwise). Color indicates infrastructure partner: Akamai in blue, Fastly in red, Cloudflare did not have any moved IP prefixes.}
        \label{fig:moved_ip_map}
    \end{minipage}
\end{figure}

For qualitative purposes, we plot the movement of IP prefixes in the egress
dataset in Figure~\ref{fig:moved_ip_map}. The color of the arcs in the figure
indicate the CDN that owns the prefix: Akamai moves are blue, Fastly are red,
Cloudflare did not move any prefixes during the study. The arcs also indicate
direction of the move as they can be followed clockwise, for instance, a large
number of Fastly prefixes moved from the west coast of the United States to
Florida. We see that prefixes associated with Fastly are most commonly moved,
with Akamai limited to relatively few. We also see that moves are not
constrained to small geographic distances, rather they are commonly moved
thousands of miles. The median change is 2,553.9 miles, 25th percentile is
605.8, and 75th percentile is 5,599.4 miles. Further, prefix moves are not
limited to within a single country. This map illustrates the dramatic geographic
changes to the ground truth dataset, changes that IP geolocation services must
quickly respond to by updating their databases.

While IP prefixes can and do change location or ownership between autonomous
systems in all networks including residential or mobile ISPs along with cloud
providers, we believe the nature and goals of Private Relay presents a uniquely
difficult challenge for IP geolocation. Cloud infrastructure is characterized by
its flexibility and dynamism---cloud IPs and services are ephemeral by their
nature and can be brought online or offline in minutes. This paradigm, which is
a fundamental reason for the dominance of cloud infrastructure, can be
detrimental for services that perform best for relatively static information
such as IP geolocation services. Overall, the results show that IP geolocation
services are facing a new and difficult challenge for Private Relay users:
accurately locating end users using the lense of cloud infrastructure IPs, which
can undergo unpredictable changes.

\subsubsection{IP geolocation service evolution}\label{sec:mm_over_time}
Instability in the advertised egress IP information has a direct effect on IP
geolocation service accuracy, as any lag in the database update process could
result in inaccurate results for some time. We can leverage our dataset to
discover geolocation service updates over time with respect to newly added IPs
to the system. We parse the duration of the egress IP dataset to identify IP
prefixes that were added during the measurement campaign.

\begin{figure}[t]
    \centering
    \begin{minipage}{.9\columnwidth}
        \includegraphics[width=1\columnwidth]{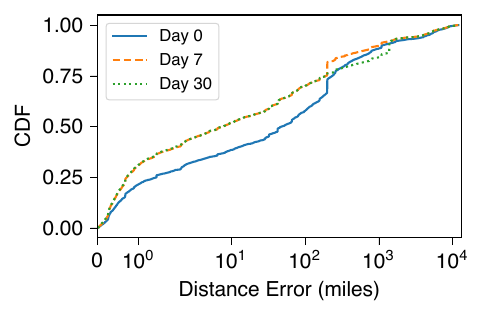}
        \caption{MaxMind geolocation accuracy for newly advertised IP addresses over time.}
        \Description{MaxMind geolocation accuracy for newly advertised IP addresses over time.}
        \label{fig:mm_over_time}
    \end{minipage}
\end{figure}

To explore the evolution of geolocation service accuracy over time, we
only select newly added IP prefixes that remained in the dataset for at least 30
days and were advertised to be in a single location throughout, resulting in a
total of 7,405 prefixes. We plot geolocation error CDFs for MaxMind in Figure~\ref{fig:mm_over_time}. We plot error distributions
for added IP prefixes at day 0, 7 and 30. As shown, the poorest accuracy tends
to occur for prefixes on day zero when they first appear in the egress dataset.
This result can be expected, as the MaxMind database is updated twice per week
and newly added IPs may not have been previously present in the database or
could be associated with erroneous past geographic information until MaxMind has
the chance to update its database. As expected, we see on day 7 the accuracy is
improved over the day zero case, with a median value of 7.8 miles versus an
initial median of 48.3. Interestingly, day 30 results largely match day 7,
leading to the conclusion that MaxMind does not appear to refine its database
beyond any early updates as new IPs are added to the PR service. Unexpectedly,
we observe a small number of cases where day 30 is less accurate than day 7. We
are unable to explain this result. Overall, it appears that users of Private
Relay can expect slightly poorer IP geolocation as geolocation services update
their databases based on new information. We also observe that the overall
distribution for newly added prefixes is rather broad, with a significant
portion (\eg, 25\%) resulting in distance errors greater than 190 miles.

\section{Discussion}\label{sec:discussion} Multi-party relay architectures like
Apple's PR are designed to achieve a compromise between performance and privacy.
However, while doing so, PR renders IP geolocation services ineffective in
determining the physical location of clients.  The previous section sheds light
on how IP geolocation services are affected by the use of MPR architectures like
Apple's iCloud Private Relay. In this section we discuss the implications of our
findings and outline potential paths forward.

\paragraph{What is the role of IP geolocation in the modern Internet?}
IP geolocation is a key element of how the Internet functions today and its role
is poised to increase in importance. As more applications aim to reduce latency
to deploy real-time services, the ability to localize users at fine-grained
levels could greatly impact the ability of such services to succeed. This raises
questions on the fundamental role of IP geolocation moving forward. In the
modern Internet, IP geolocation services serve, through a single tool, two main
purposes: mapping both clients and infrastructure to a physical location. This
approach has inherent problems as the two use cases have different requirements.
For example, the location of infrastructure is (most often) not a privacy
concern, while the location of clients may be.

Overall, the concept of IP geolocation is inherently problematic when applied to
user localization. A simple example of the problem is roaming in cellular
networks, where users remain connected to a home gateway located in the country
of origin while be physically in a different location. VPNs have abused this
concept for years, allowing users to appear as if they were in a different
country and inciting battles between content providers and
VPNs~\cite{khan2023stranger}. Our results demonstrate that the adoption of
multi-party relay architectures, such as PR, will only exacerbate the issue.

To combat this problem, we argue that we should fundamentally rethink how modern
IP geolocation operates. We believe that to achieve both aforementioned
applications---\ie, infrastructure and users localization, we need to decouple
the two to allow for different requirements and intents.

\paragraph{How to localize infrastructure?} Identifying the location of
infrastructure is a problem that has been studied for years
(\eg,~\cite{eriksson2010learning,gueye2004constraint,youn2009statistical}). The
most common approach requires the use of active tools such as traceroute and
ping to identify the location of a network interface. However, this approach is
not always accurate as the last hop responding in a traceroute experiment may
not be the closest to the target machine. In addition, traceroute is not always
available, especially in the case of IPv6, meaning there is space for
improvement.

To avoid relying on these techniques with PR, Apple opts to provide a
hand-crafted map of the IP addresses belonging to the PR infrastructure to their
nominal location. The results in the previous sections presented the issues of
such approach: the hand-crafted map is not always complete and accurate and can
easily generate inconsistencies with the results provided by commonly used IP
geolocation services. The viability for this approach would be to exclusively
rely on Apple (or any one central authority in an MPR architecture) for
correctness and completeness, which raises questions in terms of trust.

We believe that the best course of action would be to improve on existing
techniques and avoid relying on single organization to provide ground truth.
While providing a solution to this problem is out of the scope of this paper,
new promising techniques have emerged in recent years that could be leveraged to
solve this problem. For example, concepts from Riemannian geometry have been
recently proposed to compensate for probe inaccuracies in traceroute-based
measurement techniques~\cite{salamatian2022curvature}.

\paragraph{How to localize users without compromising their privacy?}
This study demonstrates that relying on IPs to identifies users' location is
becoming increasingly impractical, if not impossible. This raises the question
of how to obtain information on a user location without compromising their
privacy. While the solution to this problem could take many forms, for near-term
deployability reasons we envision the development of a trusted third party
architecture that could enable users to prove their location to services without
revealing their identity. Such an architecture could be built around short-lived
location tokens that could be used to prove a user's location through a
consensus based approach across multiple parties. Of course, a number of
practical challenges remain open in order to develop such an architecture. For
example, how to verify client location? How to prevent abuse of the system? How
to prevent a black market for location tokens? How to prevent a single party
from being able to track a user's location?
\section{Related Work}\label{sec:related}

\paragraph{Privacy-preserving Internet communications.}
Systems designed to provide anonymous communication have been studied for
decades, with the first well known architecture being Chaum's
mixnets~\cite{chaum1981untraceable}, which introduced the concept of multi-hop
relaying to provide anonymity. Mixnets provide multiple forms of privacy,
including sender and receiver anonymity. The fundamental design of mixnets were
later modified by Syverson \etal{} for use in real-time Internet communications
in their work on Onion Routing~\cite{syverson1997anonymous}, and subsequently
improved in the Tor system~\cite{dingledine2004tor}. All of these systems
provide privacy through the technique of
decoupling~\cite{schmitt2022decoupling}. Many systems leverage decoupling to
achieve anonymity for use cases beyond real-time communications. Chaum's blind
signatures provide a means for anonymous access and
authentication~\cite{chaum1983blind, chaum1984blind}, and were later adapted by
Privacy Pass~\cite{privacypass,ietf-privacy-pass-03}. Prior work has also directly studied Apple's iCloud
Private Relay, focusing on infrastructure usage and separation of
trust~\cite{sattler2022towards}, Internet traffic
performance~\cite{measuringpr}, and resistance to traffic analysis
attacks~\cite{pr_trafficanalysis}.

\paragraph{Privacy-preserving Internet naming}
There has been prior work implementing
privacy preservation in the global DNS system. Two prominent examples are
DNS-over-TLS (DoT) and DNS-over-HTTPS
(DoH)~\cite{hoffman2018dns,hu2016specification}. In both cases, a client sends
DNS queries to the resolver over an encrypted transport (TLS), which relies on
the Transmission Control Protocol (TCP). ODNS~\cite{odns} extends these designs
by obfuscating users queries making it impossible for DNS resolver to match
requests to a given user. EncDNS is functionally similar to
ODNS~\cite{herrmann2014encdns}; however, EncDNS doesn't address key
distribution. Most prior proposed DNS privacy mechanisms are protecting against
an adversary, but not a DNS operator. Castillo-Perez and Garcia-Alfaro evaluate
privacy-preserving DNS mechanisms, but show that they need additional measures
to enhance their security~\cite{castillo2009evaluation}.  Similarly, Query Name
Minimization is a proposal that limits what name servers see in DNS queries, but
a recursive resolver's operator still learns the domain requested and the
corresponding client who requested the domain~\cite{RFC7816}. Researchers have
also pointed out how aspects of current (operational) DNS, such as prefetching,
have privacy implications~\cite{krishnan2010dns,shulman2014pretty}. Federrath
\etal introduced a DNS anonymity service that employs broadcasting popular
hostnames and low-latency mixes for requesting less popular domains.

\paragraph{IP geolocation.} Commercial IP geolocation services often are delivered via database, and many
prior works have focused on studying the performance of these
databases~\cite{2007_imprecision_of_block_geolocation,
  2011_poese_uhlig_ip_geolocation_unreliable, 2017_gharaibeh_router_geolocation}.
Siwpersad \etal{}~\cite{siwpersad2008assessing} conducted a study to investigate
the level of geographic accuracy of two geolocation services also studied in
this work, MaxMind GeoLite~\cite{mm} and Hexasoft (IP2Location)~\cite{ip2l},
measuring the distance between the locations provided by each service. Their
findings showed that for 50\% of the addresses, the difference in distance was
less than 100km. Similarly, Huffaker checked the accuracy of geolocation
services by comparing the database results to active measurements collected by
PlanetLab nodes, discovering that 90\% of the IP addresses provided by these
services were different from the actual locations measured by the
nodes~\cite{huffaker2011geocompare}. Techniques beyond database have also been
explored, often using network
latency~\cite{2006_gueye_constraint_based_geolocation,2006_katz-basett_geolocation_delay_and_topology}.

In 2010, Shavitt \textit{et al}.~\cite{shavitt2010study} performed research to
evaluate the accuracy of several geolocation services using the DIMES Project's
Points of Presence (PoP) level map. The authors used an algorithm, which is
based on interface graphs, to assign IP addresses to PoPs on a map. They made
the assumption that if IP addresses are in the same PoP, they should correspond
to the same physical location on the map. The evaluation showed that MaxMind
GeoIP, GeoBytes, and Digital Envoy had the highest accuracy in placing IP
addresses within 1 km of a given PoP, with percentages ranging from 74\% to
82\%. HostIP had a somewhat lower accuracy of 57\%. In a later study,
\citeauthor{2011_shavitt_geolocation_study} found distributions ranging from
hundreds to thousands of kilometers~\cite{2011_shavitt_geolocation_study}.
Lastly, \citeauthor{2011_poese_uhlig_ip_geolocation_unreliable} found median
accuracies to fall between tens to hundreds of kilometers for MaxMind and
IP2Location~\cite{2011_poese_uhlig_ip_geolocation_unreliable}.

Most pertinent to this work, researchers have studied IP geolocation accuracy in
the context of physical network
infrastructure~\cite{2001_padmanabhan_geoping_geocluster}. Lastly, others have
studied assumptions based on co-location of IP
prefixes~\cite{2007_imprecision_of_block_geolocation,2016_gharaibeh_geo_ip_colocality}.
This work seeks to explore IP geolocation services in the specific use case where a
provider is explicitly and publicly advertising ``ground truth'' location rather
than making location of IPs hosted on large infrastructure providers somewhat
ambiguous or up for interpretation.
\section{Conclusion}\label{sec:conclusion} IP geolocation is fundamental to how
the Internet functions today. Many applications benefit from using IP
geolocation to determine the geographic location of hosts on the Internet and
optimize their service, from search engines to content providers. Multi-party
relay architectures like Apple's iCloud Private Relay (PR) are designed to
achieve a good compromise between performance and privacy. However, PR adversely
affects the capability of IP geolocation services to accurately determine the
physical location of clients. In this study we explore the accuracy of IP
geolocation services with regard to PR by comparing their location data with
Apple's data. We observe that median errors can differ by up to 1,000 miles for
IPv4 addresses. 

Our investigation has the goal to highlight the need for the networking
community to rethink IP geolocalization from the ground up. Looking ahead, we
believe that a new systems approaches will be required to enable applications to
continue to rely on geographical information while simultaneously respecting the
increased expectation of user privacy on the Internet. We believe that the
networking community is best positioned to take the lead in this effort, and we
hope that our work will serve as a starting point for new research directions
that seek to balance user privacy and system performance.

\bibliographystyle{ACM-Reference-Format}
\bibliography{paper}

\end{document}